\documentclass{article}
\usepackage{epsfig}
\title{Modifying quantum walks: A scattering theory approach}
\author{Edgar Feldman \\ Department of Mathematics \\ Graduate Center of CUNY \\ 365 Fifth
Avenue \\ New York, NY 10016 \and Mark Hillery \\ Department of Physics \\ Hunter College
of CUNY \\ New York, NY 10021}
\begin{document}
\maketitle
\begin{abstract}
We show how to construct discrete-time quantum walks on directed, Eulerian graphs.  These
graphs have tails on which the particle making the walk propagates freely, and this makes
it possible to analyze the walks in terms of scattering theory.  The probability of entering a
graph from one tail and leaving from another can be found from the scattering matrix of
the graph.  We show how the scattering matrix of a graph that is an automorphic image of
the original is related to the scattering matrix of the original graph, and we show how the
scattering matrix of the reverse graph is related to that of the original graph.  Modifications
of graphs and the effects of these modifications are then considered.  In particular we 
show how the scattering matrix of a graph is changed if we remove two tails and replace them
with an edge or cut an edge and add two tails.  This allows us to combine graphs, that is
if we connect two graphs we can construct the scattering matrix of the combined graph 
from those of its parts.  Finally, using these techniques, we show how two graphs can be 
compared by constructing a larger larger graph in which the two original graphs are in
parallel, and performing a quantum walk on the larger graph.  This is a kind of quantum
walk interferometry,
\end{abstract}

\section{Introduction}
Quantum walks are quantum versions of random walks.  In both, a
particle on a graph moves through the graph as time progresses.  In
a classical random walk, the path that the particle takes at a given
time is determined by probabilities, while in a quantum walk it is
governed by probability amplitudes.  The result is that in a
classical random walk, the motion is diffusive, while in a quantum
walk, the motion is more akin to wave propagation.  These walks were
first proposed and studied by Aharanov, Davidovich, and Zagury
\cite{aharanov}.  They were later rediscovered by a number or
workers who were interested in them as possible sources of quantum
algorithms \cite{farhi,watrous,vazirani}.  The search for walk-based
algorithms has been successful, and search algorithms \cite{shenvi},
subset-finding algorithms \cite{ambainis1,childs1}, a quantum
algorithm that can solve a particular oracle problem exponentially
faster than is possible with any classical algorithm \cite{childs2},
and, most recently, a quantum algorithm for evaluating the NAND tree
\cite{farhi2,childs3}, have been found.  There has now been
considerable work on the properties of quantum walks, and some of it
is summarized in two relatively recent reviews \cite{kempe,kendon}.

Quantum walks come in two varieties, discrete and continuous.  Here we shall consider only
discrete walks.  In these walks, there is a unitary operator that advances the walk one time
step.  In most versions of discrete-time walks, the particle making the walk is located on the
the vertices of the graph, and states corresponding to the particle being located at a particular
vertex form an orthonormal basis for a Hilbert space $\mathcal{H}_{v}$, whose states describe
the location of the particle.  In order to guarantee the unitarity of the time-step operator, it is
necessary to enlarge the Hilbert space by adding a quantum coin.  For example, if the walk is
taking place on the line, the coin space, $\mathcal{H}_{c}$ is two dimensional.  It is spanned
by the orthonormal basis $\{ |R\rangle ,|L\rangle \}$, and if the coin is in the state $|R\rangle$
the particle moves to the right on its next step, and if the coin is in the state $|L\rangle$ it moves
to the left.  On a more complicated regular graph, the dimension of the coin space is larger,
and it is spanned by an orthonormal basis, each of whose elements corresponds to a direction.
The quantum walk takes place on the space $\mathcal{H}_{v}\otimes \mathcal{H}_{c}$.

Here we shall consider a discrete-time quantum walk in which the particle is located on the
directed edges, rather than the vertices, of the graph.  The properties of a class of walks of
 this type, in which for each directed edge going from a vertex $v_{1}$ to vertex $v_{2}$, there
is a corresponding edge going from $v_{2}$ to $v_{1}$, were explored
in \cite{hillery} and \cite{feldman}.  They have the advantage that
a coin space is unnecessary, and that it is simple to define them
for any graph.  In this case, the Hilbert space that describes the
walk is spanned by an orthonormal basis whose elements correspond to
directed edges. That is, there are two orthogonal states
corresponding to to each edge; one corresponding to the particle
being on the edge going in one direction, and the other to the
particle being on the same edge but going in the opposite direction.
In \cite{feldman} we considered walks of this type on a general
graph, $G$, connected to two tails.  Each tail is a half line
consisting of an infinite number of edges, with one end going off to
infinity and the other attached to a vertex of $G$.  The particle
propagates freely on the tails, for example, if it is on one edge
moving to the right at one step, after the next step it is on the
edge to the right of the one it was on and still moving to the
right.  The motion of the particle in the graph $G$ is more
complicated.  This arrangement allowed us to study quantum walks
from the point of view of scattering theory. Scattering theory was
first applied to quantum walks by Farhi and Gutman, in the case of
continuous-time quantum walks \cite{farhi}; it our case it is
applied to discrete-time walks. A freely moving particle approaches
$G$ on one tail, scatters in $G$, and has some amplitude to be
reflected from $G$ back onto the tail from which it came and another
amplitude to be transmitted through $G$ onto the other tail.  The
properties of a walk in which a particle starts on one tail and is
later measured to be on the other one can be found from a
transmission function that is characteristic of the graph $G$.
There is a corresponding reflection function that describes walks
that begin and end on the same tail.  Both are functions of a
complex variable and are analytic in a region including the unit
disc.

Here we would like to extend that work.  We shall first consider graphs with directed edges
in which there is not necessarily a directed edge from $v_{2}$ to $v_{1}$ if there is one from
 $v_{1}$ to $v_{2}$.  The graphs will, however be Eulerian, that is each vertex will have the
same number of edges entering it as leaving it.  In addition, we shall allow an arbitrary number
of tails.  This leads to a transmission matrix instead of a transmission function.  We shall then
consider how the transmission matrix of the graph changes when the underlying graph changes.
In particular, we shall see what happens when two tails are discarded and replaced by an
edge connecting the two vertices to which they were attached or an edge is cut and replaced
by two tails.  The transmission matrix of the
new graph can be calculated from the transmission matrix of the original graph.  We shall also
be able to combine graphs.  In particular, if we have two graphs with tails, we can remove
two tails, one from each graph, and replace them by a single edge that connects the vertices
to which they were attached, thereby connecting the two graphs.  The transmission matrix for
the combined graph can be expressed in terms of the transmission matrices of the two original
graphs.  Finally, we shall show how two graphs can be compared by constructing a larger
graph from them in which the two original graphs are in parallel.  This allows us to do a kind of
interferometry on graphs.

This approach has the advantage that it allows us to construct quantum walks on larger
graphs from walks on smaller ones.  A quantum walk is characterized by the transmission
matrix of that graph.  What we show how to do is compute the transmission matrix of a
larger graph from those of smaller graphs that are its constituents.

\section{Basic formalism}
We shall begin by defining a graph in a rather general way.  A graph
$G$ consists of a set $V$ of vertices and a set $E$ of directed
edges, and two maps, $i:E\rightarrow V$ and $t:E\rightarrow V$.
These maps associate to each edge, $e$ a point $i(e)$, which we
shall call the initial point, and a point $t(e)$, which we shall
call the terminal or end point.  We allow both loop edges in which
$i(e)=t(e)$, and multiple edges, that is distinct edges with the
same initial and terminal points. This type of ensemble is often
called a digraph in the literature; all our graphs will be digraphs
so we will drop the "di". This abstract definition has the obvious
geometric realization in which we first embed the vertices as points
in Euclidean three space, and embed the edges, each of which is a
distinct copy $e=[0_e,1_e]$ of the unit interval directed from $0_e$
to $1_e$  by mapping $0_e$ onto $i(e)$ and $1_e$ onto $t(e)$.  In the
geometric realization an edge becomes a path joining $i(e)$ to $t(e)$
and inherits its orientation from the unit interval.  Thus loops
carry an unambiguous orientation.For each $v\in V$ let $\omega_{v}=
\{ e| t(e)=v \}$ and $\tau_{v} = \{ e| i(e)=v \}$. These are,
respectively, the sets of incoming and outgoing edges at $v$.  Note
that the sets $\omega_{v_{1}}$ and $\omega_{v_{2}}$ are disjoint for
$v_{1}\neq v_{2}$, as are the sets $\tau_{v_{1}}$ and
$\tau_{v_{2}}$.  We also have that $E= \bigcup_{v\in V} \omega_{v} =
\bigcup_{v\in V} \tau_{v}$.

We shall be interested in graphs that satisfy the condition
$|\omega_{v}|=|\tau_{v}|$. Graphs with this property are called
Eulerian. If a graph is to be the underlying graph for the quantum
walks we wish to study, it must be Eulerian.

The basic picture of our quantum walk is the following.  The particle making the walk is located
 on the edges of the graph, and when it passes through a vertex it scatters.
 If the particle is on an edge between the vertices $v_{1}$ and $v_{2}$, it can either be
going from  $v_{1}$ to $v_{2}$ (corresponding to the directed edge with $i(e)=v_{1}$ and
$t(e)=v_{2}$), or it can be going from $v_{2}$ to $v_{1}$ (corresponding to the directed edge
with $i(e)=v_{2}$ and $t(e)=v_{1}$).  If it is going from  $v_{1}$ to $v_{2}$, the next time step
will carry it through $v_{2}$ onto one of the edges leaving $v_{2}$.  In order to define the walk
we need a Hilbert space that describes a particle on the directed edges of the graph, and a
unitary operator that advances the walk one time step.  This operator is constructed from
operators that describe the scattering at the individual vertices.

We first construct the Hilbert space for the quantum states of a
particle moving on the graph. Let $\Omega_{v}$ and $T_{v}$ be the
Hilbert spaces generated by taking the elements of $\omega_{v}$  and
$\tau_{v}$, respectively, as orthonormal basis elements.  Let
$U_{v}: \Omega_{v} \rightarrow T_{v}$ be the local scattering
operator, and we assume that $U_{v}$ is an isometry.  By combining
these local operators we are able to construct a unitary operator
that advances the quantum walk one step.  In particular, we define
$U: L^{2}(E) \rightarrow L^{2}(E)$, where $L^{2}(E)=\bigoplus_{v\in
V}\Omega_{v} = \bigoplus_{v\in V}T_{v}$, such that
$U|\Omega_{v}=U_{v}$. We call such a unitary $U$ a quantum structure
on the Eulerian graph $G$.

Let $G=(V,E,i,t)$ be a graph.   The reverse graph
$G_R=(V_R,E_R,i_R,t_R)$ is a graph where the set $V_R=V$ and
$E_R=E$, while $i_R (e)=t(e)$ and $t_R(e)=i(e)$. In this new graph
$G_R$ we have reversed the orientation on all the edges of the  the
geometric realization of G. We see that $(\tau_R)_v=\omega_v$ and
$(\omega_R)_v=\tau_v$. If the original graph G is Eulerian then so
is $G_R$ . If $U$ is a quantum structure on $G$ we will now define a
reverse quantum structure $U_R$ on $G_R$. Let $L^{2}(E_{R})$ be
the Hilbert space generated by taking the oriented edges of $G_{R}$ as orthonormal
basis elements.  Define the conjugate linear map $R:\ L^{2}(E_{R})\rightarrow
L^{2}(E)$ so that if $|v_{2},v_{1}\rangle \in L^{2}(E_{R})$ is the basis
element corresponding to the edge $e$ considered as an edge of $G_{R}$, so
that $i_{R}(e)= v_{2}$ and $t_{R}(e)=v_{1}$, then
\begin{equation}
R|v_{2},v_{1}\rangle = |v_{1},v_{2}\rangle ,
\end{equation}
where $|v_{1},v_{2}\rangle$ is the basis element corresponding to $e$ considered
as an edge of $G$, so that $i(e)= v_{1}$ and $t(e)=v_{2}$.  $R$ is the orientation
reversing map, and it is an isometry from $(\Omega_R)_v$ onto $T_v$ and from
 $(T_R)_v$ onto $\Omega_v$, for all $v\in V$.  The reverse quantum structure on
$G_R$ is defined by
\begin{equation}
(U_R)_{v}=R^{-1}U^{-1}R ,
\end{equation}
which is a linear isometry from $(\Omega_R)_v$ onto $(T_R)_v$.
The operator $R$ defined here is closely related to the time-reversal
operator for quantum walks defined in Ref.\ \cite{feldman}.

A graph $G=(V,E,i,t)$ possesses a pairing if there exists a 
fixed-point-free involution $A$  on the edges E such that $t(Ae)=i(e)$ and
$i(Ae)=t(e)$. The initial and terminal points of $e$ are the
terminal and initial points of $Ae$. A graph possessing a pairing is
clearly Eulerian. The simplest case of such a graph is when each
pair of vertices which are connected at all have exactly two edges
joining them one in each direction, a divided highway. Such graphs
are said to be simple.  Quantum walks on simple graphs were treated
in \cite{feldman}.

Let $G=(V,E,i,t)$ and $G'=(V',E',i',t')$ be a pair of graphs. Let
$\phi=(f,F)$ be a pair of maps, $f:V \rightarrow V'$ and
$F:E\rightarrow E'$. $\phi$ is a graph morphism if $i' \circ F=f\circ i$
and $t' \circ F=f\circ t$. If
$f$ and $F$ are bijections we call $\phi$ a graph isomorphism. If $\phi$
is a graph morphism then $F:\omega_v\rightarrow \omega_{f(v)}$ and
$F:\tau_v\rightarrow\tau_{f(v)}$. $F$ thereby extends to a linear map
also denoted by $F$ which maps $\Omega_v$ into $\Omega_{f(v)}$ and
also maps $T_v$ into $T_{F(v)}$. Clearly $F:L^{2}(E)\rightarrow
L^{2}(E')$ and $\|F\|\leq1$. Let $G$ have a quantum structure $U$ and
let $G'$ have a quantum structure $U'$. A graph morphism $\phi$  is said
to be a quantum graph morphism if
\begin{equation}
F\circ U=U'\circ F .
\end{equation}
If $\phi$ is a graph isomorphism or a graph automorphism which commutes with
quantum structures as above then we call them quantum isomorphisms
or quantum automorphisms respectively.

In order to apply scattering theory to our walk, we need regions
where the particle propagates freely, and no scattering takes place.
For this reason, we will attach semi-infinite lines, or tails, to
our graph.  Let $G$ be a graph with a finite number of edges and
vertices, and $\mathcal{H}_{G}$ be the Hilbert space spanned by the
states corresponding to its directed edges.  We shall single out two
subsets of the vertices, $\{ v_{k}| k=1, \ldots K\}$, and $\{u_{l}|
 l=1, \ldots L\}$  where we will attach incoming tails $X_{k}$ and
outgoing tails $Y_{l}$ respectively.  We shall make no further
assumptions on these vertices.  A given vertex may appear many times
in one or both lists. We want to be able to clearly identify the distinct incoming
and outgoing tails. The vertices of of the incoming tail $X_{j}$
are denoted by $k_{j}$, where $k = 1, 2,\ldots $, and the
directed edges are $|k_{j},(k-1)_{j}\rangle = |k,k-1\rangle_{j}$ for
$k \geq 2$ and the attaching edge is $|1_{j},v_{j}\rangle=
|1,0\rangle_{j}$. The vertices of the outgoing tail $Y_{l}$ are
denoted by $m_{l}$, where $m=1,2,...$ . The oriented edges are
$|m_{l},(m+1)_{l}\rangle= |m,m+1\rangle_{l}$, and the attaching edge
is $|u_{l},1_{l}\rangle= |0,1\rangle_{l}$. In this way given G and
the two sets of vertices we can construct a new graph
$\Gamma=(G,(v_{1},...,v_{K})(u_{1},...,u_{L}))$ where the vertices
of new graph are those of $G$, the vertices $k_{j}$ $1\leq k<\infty$
for each tail $X_{k}$, and the vertices $(m)_{l}$ $1\leq m< \infty$
for each outgoing tail $Y_{l}$. The edges of $\Gamma$ are the edges
of $G$ and the edges $|k,k-1\rangle_{j}$, and $|m,m+1\rangle_{l}$ as
above. In order that $\Gamma $ be the underlying graph for a quantum
structure it must be Eulerian , which implies that $K=L$, i.e.\ the number
of incoming tails is the same as the number of outgoing tails. We
call such a graph $\Gamma$ an Eulerian graph with tails.

 If we are given an Eulerian graph with tails $\Gamma$, we want to study
 quantum structures $U$ on $\Gamma$ with the additional property that
\begin{equation}
U |k+1,k\rangle_{j}= |k,k-1\rangle_{j}
\end{equation}
on each edge of an incoming tail, and
\begin{equation}
U |m-1,m\rangle_{l}=|m,m+1\rangle_{l}
\end{equation}
on each edge of an outgoing tail. We say such a quantum structure is free.
The "particle" freely propagates towards $G$ along a incoming edge
and freely propagates away from $G$ along an outgoing edge.

We now want to consider eigenstates of $U$ that correspond to the
following situation.  A particle approaches $G$ on the tail $X_{k}$,
scatters in $G$, and then has amplitudes to leave $G$ on any of
outgoing tails $Y_{l}$ .  A quantum state of this form is given by
\begin{equation}
|\psi_{k}(\theta )\rangle =
\sum_{l=0}^{\infty}e^{-il\theta}|l+1,l\rangle_{k} +|\psi_{G,k}
(\theta )\rangle +  \sum_{l=1}^{K}t^{(k)}_{l}(\theta )
\sum_{m=0}^{\infty}e^{im\theta}|m,m+1\rangle_{l} ,
\end{equation}
and satisfies the equation
\begin{equation}
U|\psi_{k}(\theta )\rangle = e^{-i\theta}|\psi_{k}(\theta )\rangle .
\end{equation}
The first part of $|\psi_{k}(\theta )\rangle$ corresponds to the
incoming particle, the second to part of the state inside $G$, and
the final part to the outgoing particle.  As we shall later show,
the functions $t^{(k)}_{l}(\theta )$ and $|\psi_{G,k}(\theta
)\rangle$ are restrictions to the unit circle of functions that are
analytic for $|z|<1+\epsilon$, for some $\epsilon >0$, while
$|\psi_{k}(\theta )\rangle$ itself is the restriction to the unit
circle of an analytic function from the punctured disc,
$0<|z|<1+\epsilon$ into $L^{\infty}(E)$.

In order to begin the extension of the above eigenstate into the complex plane, define
\begin{eqnarray}
|\sigma_{j+}(z)\rangle & = & \sum_{l=0}^{\infty} z^{l}| l, l+1\rangle_{j} , \nonumber  \\
|\sigma_{j-}(z)\rangle & = & \sum_{l=0}^{\infty} z^{-l}| l+1, l\rangle_{j}  .
\end{eqnarray}
Note that
\begin{eqnarray}
U|\sigma_{j+} (z)\rangle & = & \frac{1}{z}(|\sigma_{j+}(z)\rangle - |0,1\rangle_{j}) , \nonumber  \\
U|\sigma_{j-} (z)\rangle & = & \frac{1}{z} |\sigma_{j-}(z)\rangle + U|1,0\rangle_{j}  .
\end{eqnarray}
We then define
\begin{equation}
|\psi_{k}(z)\rangle = |\sigma_{k-}(z)\rangle + |\psi_{G,k}(z)\rangle
 + \sum_{j=1}^{K} t_{j}^{(k)}(z) |\sigma_{j+}(z)\rangle ,
\end{equation}
such that it is the solution to the equation
\begin{equation}
\label{psik} z U|\psi_{k}(z)\rangle = |\psi_{k}(z)\rangle .
\end{equation}
This equation will be satisfied if and only if
\begin{eqnarray}
\label{key}
U( |\psi_{G,k}(z)\rangle + |1,0\rangle_{k}) & = & \frac{1}{z}(|\psi_{G,k}(z)\rangle \nonumber \\
 & & + \sum_{j=1}^{K} t_{j}^{(k)}(z) |0,1\rangle_{j} ) .
\end{eqnarray}
This will be our key equation, and we shall now analyze it in more detail.

Let $P_{G}$ be the orthogonal projection onto $\mathcal{H}_{G}$, and set
\begin{equation}
|w_{k}\rangle = P_{G}U|1,0\rangle_{k}  .
\end{equation}
If we now apply $P_{G}$ to both sides of Eq.\ (\ref{key}), we have that
\begin{equation}
P_{G}U|\psi_{G,k}(z)\rangle + |w_{k}\rangle = \frac{1}{z}|\psi_{G,k}(z)\rangle .
\end{equation}
Defining $U_{G}=P_{G}U$, let us consider the equation
\begin{equation}
\label{phiz}
(-zU_{G}+I)|\Phi (z)\rangle = z|w_{k}\rangle ,
\end{equation}
on $\mathcal{H}_{G}$.  We find that
\begin{eqnarray}
|\Phi (z)\rangle & = & \sum_{n=0}^{\infty}z^{n+1}U_{G}^{n}|w_{k}\rangle \nonumber \\
 & = & \sum_{n=1}^{\infty} z^{n} U_{G}^{n}|1,0\rangle_{k}  ,
\end{eqnarray}
which converges absolutely and uniformly on the disc $|z|<1$, and hence $|\Phi (z)\rangle$
is analytic in the disc $|z|<1$.

Let $\mathcal{H}_{0}$ be the subspace of $\mathcal{H}_{G}$ spanned
by the $L^{2}$ eigenvectors of $U$ that are contained in
$\mathcal{H}_{G}$, i.e. eigenstates of $U$ that have their support
in the graph $G$.  These are the analogs of bound states in
conventional potential scattering, and in that case, the bound
states are orthogonal to the scattering states.  A similar situation
obtains here.  Let $\mathcal{H}_{1}$ be the orthogonal complement of
$\mathcal{H}_{0}$ in $\mathcal{H}_{G}$, let $P_{1}$ be the
orthogonal projection onto $\mathcal{H}_{1}$, and let
$U_{1}=P_{1}U$.  It is easily seen that $|w_{k}\rangle$ and all of
its images under $U_{1}$, are orthogonal to $\mathcal{H}_{0}$.  We
now have that $|\Phi (z)\rangle$ can be expressed as
\begin{equation}
|\Phi (z)\rangle = \sum_{n=0}^{\infty} z^{n+1}U_{1}^{n}|w_{k}\rangle ,
\end{equation}
takes its values in  $\mathcal{H}_{1}$, is unique, and is analytic for $|z|<1$.  The solution can
be extended beyond the disc $|z|<1$ by making use of the fact that the operator
$(I_{\mathcal{H}_{1}}-zU_{1})$ on $\mathcal{H}_{1}$ has no eigenstates in the closed disc
$|z|\leq 1$.  Because  this is an operator on a finite dimensional space, it has a finite number
of eigenvalues, one of which is closest to $z=0$.  Let the magnitude of this eigenvalue be $r$.
For $|z|<r$, the inverse of $(I_{\mathcal{H}_{1}}-zU_{1})$ on $\mathcal{H}_{1}$  exists and is
analytic \cite{kato}, and therefore, for $|z|<r$ we have that $|\Phi (z)\rangle$ exists and is
analytic.

Summarizing, we have that Eq. (\ref{phiz}) has a unique solution, $|\Phi (z)\rangle =
|\psi_{G,k}\rangle$, with values in $\mathcal{H}_{1}$, which is analytic on a domain
$|z|<1+ \epsilon$, for some $\epsilon >0$.  Furthermore, we have that
\begin{equation}
 t_{j}^{(k)}(z) = \,_{j}\langle 0,1|U( |\psi_{G,k}(z)\rangle + |1,0\rangle_{k}) ,
\end{equation}
so that $t_{j}^{(k)}(z)$ is analytic for $|z|<1+\epsilon$, because
$|\psi_{G,k}(z)\rangle$ is. We also note that $t^{(k)}_{j}(0)=0$.

\section{Spectral results}
Let $\mathcal{H}_{k}$ be the closed, linear, $U$-invariant subspace of $\mathcal{H}=L^{2}(E)$
spanned by the states $U^{l}|j,j-1\rangle_{k}$, for $j=1,2\ldots$ and $l$ an integer.  This is just
the space of states generated by incoming states on the $k^{\rm th}$ tail.  We want to
construct the spectral representation of $U$ on $\mathcal{H}_{k}$.  We have that
\begin{enumerate}
\item $U|\psi_{k}(e^{i\theta})\rangle = e^{-i\theta}|\psi_{k}(e^{i\theta})\rangle $
\item $\langle \psi_{k}(e^{i\theta})|1,0\rangle_{k} =1$
\item $ \langle \psi_{k}(e^{i\theta})|v\rangle =0 $ for any $|v\rangle \in \mathcal{H}_{0}$ .
\end{enumerate}
If $|v\rangle \in L^{1}(E)$, then  $ \langle \psi_{k}(e^{i\theta})|v\rangle$ is a continuous function
of $\theta$, so
\begin{equation}
\frac{1}{2\pi} \int_{0}^{2\pi} d\theta |\psi_{k}(e^{i\theta})\rangle\langle \psi_{k}(e^{i\theta})|v\rangle
\in L^{\infty}(E) .
\end{equation}
The arguments in \cite{feldman} then show that for any $|y\rangle \in \mathcal{H}_{k}$ and for
$f$ any complex-valued continuous function on the unit circle (which we shall denote
by $C$), that
\begin{equation}
\langle y|f(U)|y\rangle = \frac{1}{2\pi} \int_{0}^{2\pi} d\theta f(e^{-i\theta})|\langle \psi_{k}(\theta )|
y\rangle |^{2}  .
\end{equation}
Therefore, by the Riesz-Markov theorem, there exists a unique measure, $\mu_{y}$, a Borel
measure on $C$, called the spectral measure associated to $|y\rangle$, such that
\begin{equation}
 \langle y|f(U)|y\rangle = \int_{0}^{2\pi} d\mu_{y}(\theta ) f(e^{-i\theta }) ,
\end{equation}
and hence
\begin{equation}
d\mu_{y} =\frac{|\langle \psi_{k}(\theta )|y\rangle |^{2} }{2\pi} d\theta
\end{equation}

Now define the operator $V_{k}: \mathcal{H}_{k} \rightarrow L^{2}(C)$ by
\begin{equation}
V_{k}|v\rangle = \frac{1}{\sqrt{2/\pi}} \langle\psi (e^{i\theta})|v\rangle .
\end{equation}
We then have the following theorem \cite{feldman}: $V_{k}$ is a unitary operator from
$\mathcal{H}_{k}$ to $L^{2}(C)$ such that for any $g(\theta )\in L^{2}(C)$,
we have that
\begin{equation}
V_{k}UV_{k}^{-1}g(\theta )=e^{-i\theta}g(\theta ) .
\end{equation}
Hence, we have constructed part of the spectral decomposition of $U$.  The $U$-invariant
subspaces $\mathcal{H}_{k}$ are orthogonal for different values of $k$, and they are all
orthogonal to $\mathcal{H}_{0}$.  We have, in fact, that $L^{2}(E)=\mathcal{H}_{0}
\bigoplus_{k=1}^{K}\mathcal{H}_{k}$ \cite{feldman}.  This, then, completes the spectral
decomposition of $U$.

\section{Properties of the transmission coefficients}
The transmission coefficients, $t_{j}^{(k)}(\theta )$ describe the behavior of a particle that starts
on the $k^{\rm th}$ tail and scatters into the $j^{th}$ tail.  Many properties of a quantum walk
can be found directly from these functions.

Suppose we start a walk in the state $|0,1\rangle_{k}$, and after each time step we measure
the edge $|0,1\rangle_{j}$ in order to see if the particle has arrived there.  The probability that
we find the particle there after $n$ steps, but did not find it there for any of the previous $n-1$
steps, which we shall denote by $q_{j}^{(k)}(n)$, is \cite{feldman}
\begin{equation}
q_{j}^{(k)}(n) = |\,_{k}\langle 0,1|U^{n}|1,0\rangle_{j}|^{2} .
\end{equation}
This probability can be expressed in terms of $t_{j}^{(k)}(\theta )$ as follows.  We have already
seen that $|\psi_{k}(z)\rangle$ is analytic in a region including the unit disc, and an examination
of Eq.\ (\ref{psik}) shows that it vanishes when $z=0$.  This implies that the functions
$t_{j}^{(k)}(z )$ and $\psi_{G,k}(z)\rangle$ are also analytic and vanish at $z=0$.  This implies
that
\begin{equation}
|1,0\rangle_{k}=\frac{1}{2\pi i}\int_{C} dz \frac{1}{z}|\psi_{k}(z)\rangle = \frac{1}{2\pi}\int_{0}^{2\pi}
d\theta |\psi_{k} (\theta )\rangle .
\end{equation}
We then have
\begin{equation}
\,_{k}\langle 0,1|U^{n}|1,0\rangle_{j} = \frac{1}{2\pi}\int_{0}^{2\pi} d\theta e^{-in\theta}
t_{j}^{(k)}(\theta ) ,
\end{equation}
so that
\begin{equation}
q_{j}^{(k)}(n) = \left| \frac{1}{2\pi}\int_{0}^{2\pi} d\theta e^{-in\theta} t_{j}^{(k)}(\theta ) \right|^{2} .
\end{equation}
This also gives us that the probability to find the particle on the $j^{\rm th}$ tail at some step is
given by
\begin{equation}
P_{j,out}^{(k)}= \sum_{n=1}^{\infty} q_{j}^{(k)}(n) = \frac{1}{2\pi}\int_{0}^{2\pi} d\theta
| t_{j}^{(k)}(\theta )|^{2} .
\end{equation}
This probability, and other probabilities of interest, can also be expressed in terms of 
 $t^{(k)}_{j}(z)$  by means of contour integrals \cite{feldman}.

The transmission coefficients also satisfy an orthogonality
relation. In order to derive it, first define
\begin{equation}
|\phi_{k}(\theta )\rangle = |\psi_{G,k}(\theta )\rangle + |1,0\rangle_{k} .
\end{equation}
We then have that
\begin{equation}
\label{tjk1}
\langle\phi_{l}(\theta )|\phi_{k}(\theta )\rangle = \langle \psi_{G,l} (\theta )|\psi_{G,k}(\theta )\rangle
+\delta_{k,l} .
\end{equation}
If we apply $U$ to $|\phi_{k}(\theta )\rangle$, we find
\begin{equation}
U|\phi_{k}(\theta )\rangle = e^{-i\theta} [ |\psi_{G,k}(\theta )\rangle + \sum_{j=1}^{K}t_{j}^{(k)}(\theta ) |0,1\rangle_{j} ] ,
\end{equation}
from which it follows that
\begin{equation}
\label{tjk2}
\langle\phi_{l}(\theta )|\phi_{k}(\theta )\rangle = \langle \psi_{G,l} (\theta )|\psi_{G,k}(\theta )\rangle
+ \sum_{j=1}^{K} t_{j}^{(l)}(\theta )^{\ast} t_{j}^{(k)}(\theta ) .
\end{equation}
Comparing Eqs.\ (\ref{tjk1}) and (\ref{tjk2}), we see that
\begin{equation}
 \sum_{j=1}^{K} t_{j}^{(l)}(\theta )^{\ast} t_{j}^{(k)}(\theta ) = \delta_{k,l} .
\end{equation}

Let $\tau( \partial G)=(|0,1\rangle_{l}, 1\leq l\leq K)$ and
$\omega( \partial G)=(|1,0\rangle_{k}, 1\leq k\leq K)$ be the edges
pointing out of G and the edges pointing into G respectively. Let
$T(\partial G)$ and $\Omega(\partial G)$ be the corresponding
subspaces of $L^{2}$ spanned by these edges. If we now define our
scattering matrices
\begin{equation}
S(\theta):\Omega(\partial G)\rightarrow T(\partial G)
\end{equation}
by
\begin{equation}
S(|1,0\rangle_{k})= \sum_{l=1}^{K} t^{(k)}_{l}(\theta)|0,1\rangle_{l} ,
\end{equation}
our calculation shows that for each value of
$\theta$, $S(\theta)$ is an isometry.  We also see that
\begin{equation}
\frac{1}{2\pi} \int_{0}^{2\pi}|t^{(k)}_{j}(\theta)|^{2}d\theta
\end{equation}
is equal to the probability that a particle which starts at
$|1,0\rangle_{k}$ exits G at $|0,1\rangle_{j}$ into the $j^{th}$
outgoing tail $Y_{j}$.

We will now investigate how this scattering matrix behaves when we
subject $\Gamma$ to an automorphism, and how it behaves under reversal. Let
$\Phi=(f,F)$ be quantum automorphism on $\Gamma$ which induces a
permutation $\pi_{\omega}$ on the edges in $\omega(\partial G)$ and
a permutation $\pi_{\tau}$ on the edges of $\tau(\partial G))$.
These permutations actually permute the corresponding tails. It is
easy to see that
\begin{equation}
S\circ \pi_{\omega}= \pi_{\tau}\circ S  ,
\end{equation}
if we extend $\pi_{\omega}$ and $\pi_{\tau}$ to be linear isometries on
$\Omega(\partial G)$ and $T(\partial G)$ respectively. If we write
$\pi_{\omega}|1,0\rangle_{k}=|1,0\rangle_{\pi_{\omega}(k)}$ and
$\pi_{\tau}|0,1\rangle_{j}=|0,1\rangle_{\pi_{\tau}(j)}$ we see that
\begin{equation}
t^{(k)}_{j}(\theta)=t^{(\pi_{\omega}(k))}_{\pi_{\tau}(j)}(\theta) .
\end{equation}

The discussion of the effect of reversing the graph is somewhat more
complicated. First we have to extend the reversing construction to
graphs with a free quantum structure.  Let $\Gamma$ be a given
Eulerian graph with incoming (outgoing) tails $X_{1},...,X_{K}$
$(Y_{1},...Y_{K})$ attached to $G$ at vertices $v_{1},...v_{K}$
$(u_{1},...u_{K})$ respectively. Let $\Gamma^{R}$ be the reverse of
$\Gamma$. We essentially reverse the orientation on all the oriented
edges on $\Gamma$ which means that we swap the incoming edges for
outgoing ones and vice versa. More specifically let $R$ be the
reversing map we defined in section 2 which reverses the orientation
of each edge. Then $Y^{R}_{k}$ $(X^{R}_{j})$ is the $k^{th}$
$(j^{th})$ outgoing (incoming) tail of $\Gamma^{R}$, which has the
same vertices as $X_{k}$ $(Y_{j})$ but with the orientation on the
edges reversed. If $(l_{k})$, where  $l=1,2,...$ ($(m_{j})$, where
$m=1,2...$) are the vertices $X_{k}$ $(Y_{j})$ then the edges of
$Y_{k}^{(R)}$ $(X_{j}^{(R)})$ are $|(l)_{k},(l+1)_{k}\rangle=
|l,l+1\rangle_{k}^{(R)}= R^{-1}(|l+1,l\rangle_{k})$
$(|(m+1)_{j},(m)_{j}\rangle= |m+1,m\rangle_{j}^{(R)}=
R^{-1}(|m,m+1\rangle_{j})$. Let $U^{(R)}=R^{-1}U^{-1}R$ be the
induced quantum structure on $\Gamma^{(R)}$. Then
\begin{equation}
U^{(R)}|l-1,l\rangle^{(R)}_{k}=R^{-1}U^{-1}|l,l-1\rangle_{k}=R^{-1}|l+1,l\rangle_{k}
=|l,l+1\rangle^{(R)}_{k} .
\end{equation}
Similarly we can show that
\begin{equation}
U^{(R)}|m+1,m\rangle^{(R)}_{j}= |m,m-1\rangle^{(R)}_{j} .
\end{equation}
Therefore the induced quantum structure
on $\Gamma^{(R)}$ is free.

Let $\mathcal{H}_{G}$ and $\mathcal{H}^{(R)}_{G}$ be the Hilbert spaces generated by the
interior edges of $G$ and $G^{(R)}$ respectively . Let $P_{G}$ and
$P^{(R)}_{G}$ be the orthogonal projections onto $\mathcal{H}_{G}$ and
$\mathcal{H}^{(R)}_{G}$, respectively. It is easy to see that
\begin{equation}
P_{G}\circ R^= R\circ P^{(R)}_{G} \hspace{1cm} R^{-1}\circ
P_{G}=P^{(R)}_{G}\circ R^{-1}.
\end{equation}
Let $U^{(R)}_{G}=P^{(R)}_{G}\circ U^{(R)}$. We will now explicitly construct
\begin{equation}
|\psi^{(R)}_{j}(z)\rangle=|\sigma^{(R)}_{j^{-}}(z)\rangle +
|\psi^{(R)}_{G,j}(z)\rangle + \sum^{K}_{l=1}
(t^{(R)})^{(j)}_{l}(z)|\sigma^{(R)}_{l^{+}}(z)\rangle ,
\end{equation}
where
\begin{equation}
|\psi^{(R)}_{G,j}(z)\rangle= \sum_{n=1}^{\infty}z^{n}(U_{G}^{(R)})^{n}|1,0\rangle_{j}^{(R)} .
\end{equation}
formula (18) shows that
\begin{eqnarray}
(t^{(R)})_{k}^{(j)}(z) & = & \langle(0,1)^{(R)}_{k}|U^{(R)}[|\psi^{(R)}_{G,j}(z)\rangle
+|1,0\rangle^{(R)}_{j}]\rangle  \nonumber \\
 & = & \langle R(0,1)^{(R)}_{k}| R U^{(R)}[|\psi^{(R)}_{G,j}(z)\rangle
+|1,0\rangle^{(R)}_{j}]\rangle^{\ast}  \nonumber  \\
 & = & \langle U (1,0)_{k}|R[|\psi^{(R)}_{G,j}(z)\rangle] +|0,1\rangle_{j}\rangle^{\ast}.
\end{eqnarray}
Now
\begin{eqnarray}
\langle U(1,0)_{k}|R(z^{n}(U_{G}^{(R)})^{n}(1,0)_{j}^{(R)}\rangle & = & \langle
z^{n}(UP_{G})^{n}U(1,0)_{k}|(0,1)_{j}\rangle \nonumber \\
 & = & \langle z^{n}U(U_{G})^{n}(0,1)_{k}|(0,1)_{j}\rangle .
\end{eqnarray}
This implies
\begin{equation}
\label{conjana}
\langle U(0,1)_{k}|R[\psi^{(R)}_{G,j}(z)\rangle]\rangle= \langle
U\psi_{G,k}(z)|(0,1)_{j}\rangle ,
\end{equation}
which gives us the desired formula
\begin{equation}
\label{transrev}
(t^{(R)})^{(j)}_{k}(z)= \langle U(|\psi_{G,g}(z)+
(1,0)_{k})|(0,1)_{j}\rangle^{\ast} = t^{(k)}_{j}(z) .
\end{equation}
This enables us to compute the transmission amplitudes for
$\Gamma^{(R)}$ from those of our original graph $\Gamma$. If
$S(\theta)$ is the scattering matrix for $\Gamma$ then
$S^{tr}(\theta)$ is the scattering matrix $S^{(R)}(\theta )$ for
$\Gamma^{(R)}$. Furthermore we see that
\begin{equation}
R\circ S^{(R)}(\theta)\circ R^{-1}|0,1\rangle_{k}=\sum_{l=1}^{K}
t^{(l)}_{k}(\theta)^{\ast}|0,1\rangle_{l}
\end{equation}

It is because of Eq.\  (\ref{conjana}) that we need to choose $R$ to be
conjugate linear. The right hand side is clearly a conjugate
analytic function of $z$, so the fact that $R$ interchanges the complex
structures forces the left hand side to be conjugate analytic too.
When we defined the reverse graph $G_{R}$, we defined the mapping
$R:L^{2}(E_{R}) \rightarrow L^{2}(E)$ to be a conjugate linear
isometry. If instead we replace $L^{2}(E_{R})$ with$L^{2}(E_{R})_c$,
the space with the conjugate complex structure and Hermitian inner
product, $R$ becomes a complex linear isometry from $L^{2}(E_{R})_c$
onto $L^{2}(E)$. If we follow the argument above with these changes
we get the formula
\begin{equation} 
(t^{(R)})^{(j)}_{k}(z)=t^{(k)}_{j}(z)^{\ast} ,
\end{equation}
which seems to contradict analyticity until we recall that we are
dealing with $L^{2}(E_{R})_{c}$.  So the ``$z$'' on the left hand side
refers to the complex structure which is the conjugate of the
original one on $L^{2}(E)$. Hence, the more functorial formula above is as
it should be. If we followed through in this vein we would finally
arrive at Eq.\ (\ref{transrev}). We chose to avoid this additional
complication because our concern was to give formulas to compute the
transition amplitudes.

\section{Add a handle}

We now want to begin our study of how changing a graph modifies its
transmission amplitudes. Let us first consider what happens when we
replace a pair of tails, one incoming and one outgoing with a single
edge going from the attaching vertex of the outgoing edge to the
attaching vertex of the incoming edge. This new graph $\Gamma'$ will
still be Eulerian.  It will also inherit a free quantum structure $U^{\prime}$
from the original free quantum structure $U$ on $\Gamma$.

Let us be more specific.  Let
$\Gamma=(G,(v_{1},...,v_{K}),(u_{1},...,u_{K}))$ be an Eulerian
graph with tails. Let $X_{1},...X_{K}$ be the incoming tails
attached at the vertices $v_{1},...,v_{K}$ respectively
 and let $Y_{1},...,Y_{K}$ be the outgoing tails attached at the
 vertices $u_{1},...,u_{K}$ respectively. Let
 $\Gamma'=(G',(v_{2},...v_{K}),(u_{2},...,u_{K}))$ be the new Eulerian
 graph with tails. The graph $G'$ has the same vertices as $G$.  Its
 edges are those of $G$ plus a new one $|u_{1},v_{1}\rangle$, which is
 an oriented edge from $u_{1}$ to $v_{1}$. Its incoming (outgoing)
 tails are $X_{2},...X_{K}$ ($Y_{2},...,Y_{K}$) with the same
 attaching vertices as in $\Gamma$. We see that $\tau^{\prime}_{v}=\tau_{v}$
 and $\omega^{\prime}_{v}=\omega_{v}$ for vertices $v$ not equal to $u_{1}$ or
 $v_{1}$. The prime refers to $\Gamma'$. The edges of
 $\omega^{\prime}_{v_{1}}$ are the same as those of $\omega_{v_{1}}$ with
 $|1,0\rangle_{1}=|1_{1},v_{1}\rangle$ replaced by
 $|u_{1},v_{1}\rangle$. The edges $\tau^{\prime}_{u_{1}}$ are the same as
 those of $\tau_{u_{1}}$ with
 $|0,1\rangle_{1}=|u_{1},1_{1}\rangle$ replaced by
 $|u_{1},v_{1}\rangle$.  We can now see how the original free
 quantum structure $U$ on $\Gamma$ induces a new one, $U^{\prime}$, on
$\Gamma^{\prime}$.   If a vertex $v$ is unequal to $u_{1}$ or $v_{1}$ then
$U^{\prime}_{v}=U_{v}$.  If the vertex is $v_{1}$ or $u_{1}$ this is still the rule if we
 consistently replace $|1,0\rangle_{1}$ in $\omega_{v_{1}}$ with
 $|u_{1},v_{1}\rangle$ and $|0,1\rangle_{1}$ in
 $\tau_{u_{1}}$ with $|u_{1},v_{1}\rangle$. So
\begin{equation}
U^{\prime}_{v_{1}}|u_{1},v_{1}\rangle =U_{v} |1,0\rangle_{1} ,
\end{equation}
which is unambiguous unless $u_{1}=v_{1}$.  In that case $U_{v} |1,0\rangle_{1}$
 may contain a term proportional to $|u_{1},(1)_{1}\rangle$.  In defining
$U^{\prime}_{v_{1}}|u_{1},v_{1}\rangle$, we simply take the expression for
$U_{v} |1,0\rangle_{1}$ and replace $|u_{1},(1)_{1}\rangle$ by $|u_{1},v_{1}\rangle$
wherever it occurs.

Our aim, now, is to show how to simply compute the transmission
amplitudes $\tau^{(k)}_{j}(z)$ of the new configuration $\Gamma^{\prime}$ from
the transmission amplitudes $t^{(k)}_{j}(z)$ of the original $\Gamma$. We will adopt
the notation of section 2. Let
\begin{equation}
|\psi'_{k}(z)\rangle= |\sigma_{k^{-}}(z)\rangle + |\psi_{G',k}(z)\rangle +
 \Sigma_{j=2}^{K} \tau^{(k)}_{j}(z) |\sigma_{j^{+}}(z)\rangle ,
\end{equation}
be the generalized eigenstate of $U^{\prime}$ which satisfies the equation
\begin{equation}
zU^{\prime} |\psi^{\prime}_{k}(z)\rangle = |\psi^{\prime}_{k}(z)\rangle .
\end{equation}
We want to represent $|\psi_{G',k}(z)\rangle$ in the form
\begin{equation}
\label{handle1}
|\psi_{G',k}(z)\rangle= |\psi_{G,k}(z)\rangle +
 a_{k}(z)( |u_{1},v_{1}\rangle + |\psi_{G,1}(z)\rangle) ,
\end{equation}
where $|\psi_{G,k}(z)\rangle$ is that part of the generalized eigenstate
 of $\Gamma$ that is supported in $G$, and $a_{k}(z)$ is a function to
 be determined. We have that
\begin{equation}
\label{handle2}
zU^{\prime} (|u_{1},v_{1}\rangle +|\psi_{G,1}(z)\rangle)= |\psi_{G,1}(z)\rangle +
 t^{(1)}_{1}(z)|u_{1},v_{1}\rangle +
 \sum_{j=2}^{K}t^{(1)}_{j}(z)|0,1\rangle_{j} ,
\end{equation}
 and
\begin{equation}
\label{handle3}
zU^{\prime}(|1,0\rangle_{k} + |\psi_{G',k}(z))= |\psi_{G',k}(z)\rangle +
 \sum_{j=2}^{K} \tau^{(k)}_{j}(z)|0,1\rangle_{j} ,
\end{equation}
where, as previously mentioned, the $\tau^{(k)}_{j}(z)$ denote the transmission amplitudes for
 $\Gamma^{\prime}$. If we represent $|\psi_{G',k}(z)\rangle$ as in Eq.\ (\ref{handle1}, we see that
\begin{eqnarray}
zU^{\prime}( |1,0\rangle_{k} + |\psi_{G,k}(z)\rangle +
 a_{k}(z)(|u_{1},v_{1}\rangle + |\psi_{G,1}(z)\rangle)) \nonumber \\
= |\psi_{G,k}(z)\rangle + \sum_{2}^{K}t^{(k)}_{j}(z)|0,1\rangle_{j} +
 t^{(k)}_{1}(z)|(u_{1},v_{1})\rangle + a_{k}(|\psi_{G,1}(z)\rangle  \nonumber \\
+ t^{(1)}_{1}(z)|u_{1},v_{1}\rangle +\sum^{K}_{j=2}t^{(1)}_{j}(z)|0,1\rangle_{j}) .
\end{eqnarray}
Comparing this result with Eq.\ (\ref{handle3}), we can see that
\begin{eqnarray}
 \tau^{k}_{j}(z) & = & t^{(K)}_{j}(z) + a_{k}(z)t^{(1)}_{j}(z) \nonumber \\
 a_{k}(z) & = & t^{(k)}_{1}(z) + a_{k}(z)t^{(1)}_{1}(z) ,
\end{eqnarray}
so that, finally,
\begin{equation}
 \tau^{(k)}_{j}(z)  =  t^{(k)}_{j}(z) + \frac{t^{(1)}_{j}(z)\, t^{(k)}_{1}(z)} {1-t^{(1)}_{1}(z)} ,
\end{equation}
all of which hold in a neighborhood of the unit disk by the discussion in section 2.

 We can iterate this procedure by connecting an outgoing tail to an
 incoming tail as we did above, one pair at a time, thereby adding
 several "handles".  We can also use this method to accomplish this
 in a single step in the following way.  Let us try to add $L$ handles
 by splicing $Y_{j}$ to $X_{j}$ $1\leq j\leq L$ respectively and
 form new edges (handles), $e_{1}=(u_{1},v_{1})$
 $e_{2}=(u_{2},v_{2})$ ...$e_{L}=(u_{L},v_{L})$. The new generalized
 eigenstate coming from an incoming wave traveling along one of the
 remaining incoming tails $Y_{k}$ is
\begin{equation}
 |\psi'_{k}(z)\rangle=\sigma_{k^{-}}(z) + |\psi_{G',k}(z)\rangle +
 \sum^{K}_{j=L+1}\tau^{(k)}_{j}(z)|\sigma_{j^{+}}(z)\rangle ,
\end{equation}
where $\psi_{G',k}(z)\rangle$ is of the form
\begin{equation}
 |\psi_{G',k}(z)\rangle= |\psi_{G,k}(z)\rangle +
 \sum^{L}_{l=1}a_{k,l}(z)[|u_{l},v_{l}\rangle +
 |\psi_{G,l}(z)\rangle ] ,
\end{equation}
 where the functions $a_{k,l}(z)$ are to be
 determined. Exactly as before, we have that
\begin{equation}
zU'(|1,0\rangle_{k} + |\psi_{G',k}(z)\rangle)= |\psi_{G',k}(z)\rangle +
 \sum^{K}_{j=L+1}\tau^{(k)}_{j}(z)|0,1\rangle{j} ,
\end{equation}
where $\tau^{(k)}_{j}(z)$ denotes the appropriate transmission amplitude
 for $\Gamma^{\prime}$ the new Eulerian graph with tails where we replaced
 the  first $L$ pairs of tails with the corresponding handles. As before
\begin{eqnarray}
zU'(|1,0\rangle_{k} + |\psi_{G,k}(z)\rangle +
 \sum_{l=1}^{L}a_{k,l}[|1,0\rangle_{l} + |\psi_{G,l}(z)\rangle ]) \nonumber \\
= |\psi_{G,k}\rangle+ \sum_{j=L+1}^{K}t^{(k)}_{j}|0,1\rangle_{j} +
 \sum_{j=1}^{L}t^{(k)}_{j}(z)|(u_{j},v_{j})\rangle  \nonumber \\
 +\sum_{l=1}^{L}a_{k,l}(z) \left[ |\psi_{G,l}(z)\rangle +
 \sum_{j=L+1}^{K}t^{(l)}_{j}(z)|0,1\rangle_{j} +
 \sum_{j=1}^{L}t^{(l)}_{j}(z)|(u_{j},v_{j})\rangle \right]  .
\end{eqnarray}
These equations lead to two sets of equations. The first set, $L$ in number, is
\begin{equation}
 a_{k,l}(z)=t^{(k)}_{l}(z)+ \sum_{j=1}^{L}a_{k,j}(z)t^{(j)}_{l}(z) ,
\end{equation}
 where $1\leq l \leq L$. This set of equations can be solved by
 Cramer's rule to give expressions for $a_{k,l}$ as rational
 functions of $t^{(k)}_{l}(z)$ and $t^{(j)}_{l}(z)$ in some
 neighborhood of the origin in the complex plane, because,
the transmission amplitudes of $\Gamma$ vanish at the origin.
The second set of equations is
\begin{equation}
\tau^{(k)}_{j}= t^{(k)}_{j} + \sum _{l=1}^{L}a_{k,l}(z)t^{(l)}_{j}(z) ,
\end{equation}
for $L+1\leq j\leq K$.  If we
 substitute the solutions to the first set of equations into the
 second set we get our desired result, which expresses the transition
 amplitudes of $\Gamma^{\prime}$ as rational functions of the transition
 amplitudes of $\Gamma$.  The formula holds on a neighborhood of the
 unit disk in the complex plane. The rational functions themselves
 only depend upon the numbers $K$ and $L$.

 \section{Cut a handle}

 Let us start with $\Gamma$ an Eulerian graph with tails. Let
 $X_{2}$,...,$X_{K}$ ($Y_{2}$,...,$Y_{K}$) $K-1$ incoming (outgoing)
 attached at the vertices $v_{2}$,...,$v_{K}$ ($u_{2}$,...,$u_{K}$)
 respectively. Let $e=(u_{1},v_{1})$  be an oriented edge in $G$. We
 wish to replace the edge e with a pair of tails, an incoming tail
 $X_{1}$ attached at $v_{1}$ and an outgoing tail $Y_{1}$ attached
 at $u_{1}$. The new vertices on $X_{1}$ the incoming edge will be
 denoted $1_{x}$,$2_{x}$,...and the corresponding edges will be
 denoted $|k,k-1\rangle_{x}=|k_{x},(k-1)_{x}\rangle$.  The new
 vertices on $Y_{1}$ the outgoing edge will be $1_{y}$,$2_{y}$...and
 the corresponding edges will be
 $|k,k+1\rangle_{y}=|k_{y},(k+1)_{y}\rangle$ $k=1,2,...$. We will
 sometimes denote $u_{1}$ as $0_{y}$ and denote $v_{1}$ as $0_{x}$
 respectively. Let $G'$ be the graph with the same vertices as $G$ and
 the same edges as $G$ with the edge $(u_{1},v_{1})$ removed. The new
 graph with tails
 $\Gamma'=(G',(v_{1},v_{2},...,v_{K}),(u_{1},u_{2},...,u_{K}))$ has all the vertices
of $\Gamma$ as well as the new vertices $k_{x}$
 and $k_{y}$ for the two new tails. The edges of $\Gamma^{\prime}$ are those
 of $\Gamma$ with $(u_{1},v_{1})$ removed and the tail edges
 $|k,k+1\rangle_{y}$ and $|k+1,k\rangle_{x}$ $k=1,2,...$ added. The
 original free quantum structure $U$ on $\Gamma$ induces a new free
 quantum structure $U'$ on $\Gamma'$ as follows. First
 $U'_{k_{y}}|k-1,k\rangle_{y}=|k,k+1\rangle_{y}$ and
 $U'_{k_{x}}|k+1,k\rangle_{x}=|k,k-1\rangle_{x}$ must hold in order
 that $U'$ be free. If $v$ is a vertex of $G'$ which is not equal to
 $u_{1}$ or $v_{1}$ then $U'_{v}=U_{v}$. In order to deal with
 $v_{1}$ and $u_{1}$ we note as in the last section that the edges
 of $\omega^{\prime}_{v{1}}$ $(\tau^{\prime}_{u_{1}})$ are the same as
those of $\omega_{v_{1}}$ $(\tau_{u_{1}})$ with
 the edge $(u_{1},v_{1})$ of $\Gamma$ replaced by
 $|1,0\rangle_{x}=|1_{x},v_{1}\rangle$
 $(|0,1\rangle_{y}=|u_{1},1_{y}\rangle$. Thus the rule
 $U'_{v}=U_{v}$ still holds for vertices $v_{1}$ and $u_{1}$ if we
 consistently replace $|u_{1},v_{1}\rangle$ in $\omega_{v_{1}}$
 $(\tau_{u_{1}})$ by $|1,0\rangle_{x}$ $(|0,1\rangle_{y})$ so
\begin{equation}
 U'_{v_{1}}|1,0\rangle_{x} = U_{v_{1}}|u_{1},v_{1}\rangle .
\end{equation}
This is unambiguous unless $u_{1}=v_{1}$ where we replace
 $|u_{1},v_{1}\rangle$ by $|0,1\rangle_{y}$ in the formula.

 This operation is clearly the inverse of our add a handle
 procedure. If we create a pair of tails $X_{1}$ and $Y_{1}$ from an
 edge $(u_{1},v_{1})$ as above and the splice the new tails together
 to reform the edge $(u_{1},v_{1})$ according to our add a handle
 prescription we end up with the same free quantum structure on the
 same graph with tails.  Similarly if we start by first adding a
 handle by splicing a pair of tails, and then cut this new handle by
 the rules above we again arrive back where we started from.

 Let $t^{(k)}_{j}(z)$, $2\leq j,k\leq K$, be the transmission amplitudes
 for $\Gamma$, our original graph with tails, and let $T^{(k)}_{j}(z)$, $1\leq k,j\leq K$, be the
 transmission amplitudes for the new configuration $\Gamma^{\prime}$. We
 wish to compute the functions $T^{(k)}_{j}(z)$ from the attributes of $\Gamma$,
which include the functions $t^{(k)}_{j}(z)$, as simply as possible.  We begin by
applying the results of the previous section to the graph $\Gamma^{\prime}$, as
$\Gamma$ is obtained from $\Gamma^{\prime}$ by adding a handle.  This immediately
gives us that
\begin{equation}
\label{cut}
t^{(k)}_{j}(z)= T^{(k)}_{j}(z) +\frac{T^{(k)}_{1}(z)T^{(1)}_{j}(z)}{1 - T^{(1)}_{1}(z)} ,
\end{equation}
for $2\leq j,k \leq K$.  If we could find the functions $T^{(1)}_{j}(z)$ and $T^{(j)}_{1}(z)$,
for $1 \leq j \leq K$, then we could solve these equations for the remaining $T^{(k)}_{j}(z)$.
Note that these are the transmission amplitudes associated with the new tails, $X_{1}$
and $Y_{1}$.

In order to describe how to calculate these transmission amplitudes from $\Gamma$, we
define the generalized eigenstate on $\Gamma^{\prime}$
\begin{equation}
|\psi^{\prime}_{1}(z)\rangle=|\sigma_{1^{-}}(z)\rangle + |\psi_{G',1}\rangle + \sum _{j=1}^{K}T^{(1)}_{j}(z)|\sigma_{j^{+}}(z)\rangle ,
\end{equation}
This is the formula for the generalized eigenstate generated by an incoming wave
along $X_{1}$, where $|\psi_{G',1}\rangle$ is the part of this eigenstate that is supported
on $\mathcal{H}_{G'}$. Let us note that $\mathcal{H}_{G}$ is the direct sum of
$\mathcal{H}_{G'}$ and the one-dimensional subspace consisting of multiples of
$|u_{1},v_{1}\rangle$.  Let $P_{G'}$ denote the orthogonal projection operator onto
$\mathcal{H}_{G'}$, and let $U_{G'}=P_{G'}U$. Considered as a state on $\Gamma^{\prime}$,
 $|\psi_{G',1}\rangle$ satisfies the equation
\begin{equation}
zP_{G^{\prime}}U^{\prime} ( |\psi_{G',1}(z)\rangle +|1,0\rangle_{x}) =  |\psi_{G',1}(z)\rangle ,
\end{equation}
and this implies that considered as a state on $\Gamma$ it satisfies
\begin{equation}
zP_{G^{\prime}}U ( |\psi_{G',1}(z)\rangle +|u_{1},v_{1}\rangle_{x})  =  |\psi_{G',1}(z)\rangle .
\end{equation}
The solution to this equation is
\begin{equation}
 |\psi_{G',1}(z)\rangle=\sum_{n=1}^{\infty}z^{n}U_{G'}^{(n)}|u_{1},v_{1}\rangle  .
\end{equation}
Note that this equation contains only quantities defined on the original graph, $\Gamma$,
so that it implies that $|\psi_{G',1}(z)\rangle$ can be calculated from the initial graph.  Once
we have found $|\psi_{G',1}(z)\rangle$ we can substitute it into the equation
\begin{equation}
 T^{(1)}_{j}(z)=_{j}\langle0,1||U'(\psi_{G',1}(z) +|1,0\rangle_{x}) ,
\end{equation}
defined on $\Gamma^{\prime}$, to find $T^{(1)}_{j}(z)$.

Our remaining task is to find $T^{(j)}_{1}(z)$, for $2 \leq j \leq K$.  This can be done by
looking with the reverse graph $\Gamma^{R}$ and its induced quantum structure and
following a procedure analogous to the one we followed for $\Gamma$.  We first find
\begin{equation}
 |\psi_{G',1}^{R}(z)\rangle=\sum_{n=1}^{\infty}z^{n}(P^{R}_{G'}U^{(R)})^{n}|v_{1},u_{1}\rangle  ,
\end{equation}
where $P^{R}_{G^{\prime}}$ projects onto the subspace spanned by the states corresponding
to all of the edges of $G^{(R)}$ except $|v_{1},u_{1}\rangle$.  Once we have
$ |\psi_{G',1}^{R}(z)\rangle$, we can use it to find $T^{(R)(1)}_{j}(z)$ in the same we we used
$|\psi_{G',1}(z)\rangle$ to find $T^{(1)}_{j}(z)$.  Finally, we can make use of the relation
\begin{equation}
T^{(R)(1)}_{j}(z)= T^{(j)}_{1}(z),
\end{equation}
to find $T^{(j)}_{1}(z)$.  This means that all of the quantities in Eq.\ (\ref{cut}) except
$T^{(k)}_{j}(z)$ for $2 \leq j,k \leq K$ are then, in principle, known so that they can be used
to find the transmission amplitudes for the cut graph.

\section{The splice}
We can use the results of section 5 to find the transmission amplitudes of a graph
that is the result of splicing two other graphs together.  Let us assume that the graph $G$,
with entering tails $X_{1},\ldots X_{K}$ attached to $G$ at the vertices $v_{1},\ldots v_{K}$,
and exiting tails $Y_{1},\ldots Y_{K}$ attached to $G$ at the vertices $u_{1},\ldots u_{K}$,
can be broken into two disjoint pieces, $G_{1}$ and $G_{2}$.  $G_{1}$ contains the vertices
$v_{1},\ldots v_{L-1}$, to which the tails $X_{1},\ldots X_{L-1}$ are attached and the vertices
$u_{1}\ldots u_{L-1}$ to which the tails $Y_{1}\ldots Y_{L-1}$ are attached.  Similarly,
$G_{2}$ contains the vertices $v_{L},\ldots v_{K}$, to which the tails $X_{L},\ldots X_{K}$ are
attached and the vertices $u_{L}\ldots u_{K}$ to which the tails $Y_{L},\ldots Y_{K}$
are attached.  We now remove the tails $Y_{1}$ and $X_{L}$ and replace them by one
directed edge from $u_{1}$ to $v_{L}$.  This is simply a specific example of the adding a
handle construction.

Let us now apply the results of the previous section.  We first note that in the original graph, a
particle entering $G_{1}$ would never exit from $G_{2}$, so $t_{j}^{(k)}=0$ if $v_{k}\in G_{1}$
and $u_{j}\in G_{2}$.  Similarly, a particle entering $G_{2}$ would never exit from $G_{1}$,
which implies that $t_{j}^{(k)}=0$ if $v_{k}\in G_{2}$ and $u_{j}\in G_{1}$.  Making use of this
result, we have from section 5 that
\begin{equation}
\tau^{(k)}_{j}(z) = \left\{ \begin{array}{cc} t_{j}^{(L)}(z) t_{1}^{(k)}(z) &\hspace{5mm}
L\leq j \leq K,\ 1\leq k \leq L-1 \\  t_{j}^{(k)}(z) & \hspace{5mm} {\rm otherwise} \end{array} \right.
\end{equation}
Therefore, we can use the add-a-handle construction to construct a quantum walk on a larger
graph from walks on smaller ones.

\section{Comparing graphs}
We can use the techniques developed here to compare two graphs by
doing a kind of interferometry with them.  Suppose that $G_{1}$ has
only two tails, an incoming tail $X_{1}$ attached to vertex $v_{1}$
and an outgoing tail $Y_{1}$ attached to $u_{1}$.  Similarly,
$G_{2}$ also has only two tails, an incoming tail $X_{2}$ attached
to $v_{2}$ and an outgoing tail $Y_{2}$ attached to $u_{2}$.  We
would like to determine whether $G_{1}$ and $G_{2}$ are the same or
different, with the additional constraint that if they are the same,
$v_{1}$ in $G_{1}$ should correspond to $v_{2}$ in $G_{2}$, and
$u_{1}$ in $G_{1}$ should correspond to $u_{2}$ in $G_{2}$.

\begin{figure} [ht]
\epsfig{file=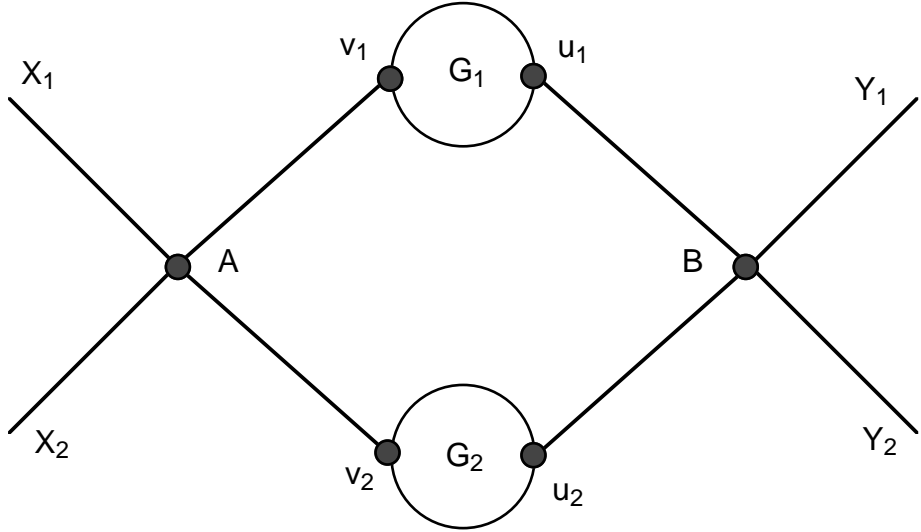}
\caption{An arrangement for comparing the graphs $G_{1}$ and $G_{2}$.
A particle starting a walk on the tail $X_{1}$ will never exit on
the tail $Y_{2}$ if the graphs $G_{1}$ and $G_{2}$ are identical.}
\end{figure}

One way of attacking this problem is to put the two graphs into an
arrangement like that shown in Figure 1.  First, we remove the tails
from the graphs.  We now consider a vertex, $A$, with two incoming
and two outgoing edges. The incoming edges are the initial edges
of two incoming tails, $X_{1A}$ and  $X_{2A}$. One outgoing edge
is attached to $v_{1}$ and the other is attached to $v_{2}$.  Similarly,
we consider a second vertex, $B$, with two incoming and two outgoing
edges.  The outgoing edges are the initial edges of two outgoing tails,
$Y_{1B}$ and $Y_{2B}$, and one incoming edge is attached to $u_{1}$
and the other is attached to $u_{2}$.  We shall denote by $\Gamma$  be the
graph with tails that we get when we connect $G_{1}$ and $G_{2}$ to the
vertices $A$ and $B$ with the tails $X_{1A}$, $X_{2A}$, $Y_{1B}$ and $Y_{2B}$,
and by $U$ the free quantum structure on $\Gamma$.

What we have done is to create a larger graph from $G_{1}$ and
$G_{2}$ in which the two graphs are in parallel.  We shall choose
the unitary operators corresponding to the vertices $A$ and $B$ in
such a way that if we start the walk on the incoming tail $X_{1A}$, and
if the graphs $G_{1}$ and $G_{2}$ are identical, with $v_{1}$ in $G_{1}$
corresponding to $v_{2}$ in $G_{2}$, and $u_{1}$ in $G_{1}$ corresponding
to $u_{2}$ in $G_{2}$, then the transmission amplitude corresponding to the
outgoing tail $Y_{2B}$ will be zero.  Therefore, if we
start a walk on the tail $X_{1A}$, and eventually find the particle on the tail
$Y_{2B}$, we can conclude that the two graphs are not identical.

We now want to find the transmission amplitudes of the combined
graph in terms of the transmission amplitudes of $G_{1}$ and
$G_{2}$.  In order to do this, we need to specify what happens at
the vertices $A$ and $B$.  The states corresponding to the edges entering
vertex $A$ are $|1_{1},A\rangle$, which is the initial edge of $X_{1A}$, and
$|1_{2},A\rangle$, which is the initial edge of $X_{2A}$.  The states corresponding
to the edges leaving vertex $A$ are $|A,v_{1}\rangle$ and $|A,v_{2}\rangle$.
The local isometry at vertex $A$ mapping incoming to outgoing states has
the following action
\begin{eqnarray}
|1_{1},A\rangle & \rightarrow & \frac{1}{\sqrt{2}}(|A,v_{1}\rangle + |A,v_{2}\rangle ) \nonumber \\
|1_{2},A\rangle & \rightarrow & \frac{1}{\sqrt{2}}(|A,v_{1}\rangle - |A,v_{2}\rangle ) .
\end{eqnarray}
The local isometry at vertex $B$ is essentially identical.  The states corresponding to
the edges entering  vertex $B$ are $|u_{1},B\rangle$ and $|u_{2},B\rangle$.  The states
corresponding to the edges leaving vertex $B$ are $|B,1_{1}\rangle$, which is the initial
edge of $Y_{1B}$, and $|B,v_{2}\rangle$, which is the initial edge of $Y_{2B}$.  The local
isometry at vertex $B$ mapping incoming to outgoing states is given by
\begin{eqnarray}
|u_{1},B\rangle & \rightarrow & \frac{1}{\sqrt{2}}(|B,1_{1}\rangle + |B,1_{2}\rangle ) \nonumber \\
|u_{2},B\rangle & \rightarrow & \frac{1}{\sqrt{2}}(|B,1_{1}\rangle - |B,1_{2}\rangle ) .
\end{eqnarray}

The generalized eigenstate corresponding to the entering particle
being on the tail $X_{1A}$ is
\begin{equation}
|\psi_{1}(z)\rangle = |\sigma_{1-}(z)\rangle + |\psi_{A,B}(z)\rangle + \sum_{j=1}^{2}\tau_{j}^{(1)}
(z)|\sigma_{j+}(z)\rangle ,
\end{equation}
where $|\psi_{A,B}(z)\rangle$ is the part between vertices $A$ and
$B$. The transmission amplitude $\tau_{1}^{(1)}(z)$ corresponds to
the tail $Y_{1B}$, and the transmission amplitude
$\tau_{2}^{(1)}(z)$ corresponds to the tail $Y_{2B}$.  We would like
to find these transmission amplitudes in terms of the transmission
amplitudes for $G_{1}$ and $G_{2}$, which we shall denote by
$t_{1,1}^{(1)}(z)$ and $t_{1,2}^{(1)}(z)$, respectively.  We can
find
 the transmission amplitudes for $\Gamma$, and the state $|\psi_{A,B}(z)\rangle$ by making use
of the equations
\begin{eqnarray}
zU( |1_{1},A\rangle + |\psi_{A,B}(z)\rangle ) & = &
|\psi_{A,B}(z)\rangle + \sum_{j=1}^{2}\tau_{j}^{(1)}(z)
|B, 1_{j}\rangle  \nonumber  \\
zU( |A,v_{k}\rangle + |\psi_{k}(z)\rangle ) & = &
|\psi_{k}(z)\rangle + t_{1,k}^{(1)}(z)
 |u_{k},B\rangle  ,
\end{eqnarray}
where $k=1,2$ in the second equation.  The states $|\psi_{k}(z)\rangle$, for $k=1,2$ are the
internal parts of the generalized eigenfunctions of the graphs $G_{1}$ and $G_{2}$,
respectively.  We find that
\begin{eqnarray}
|\psi_{A,B}\rangle & = & \frac{z}{\sqrt{2}} \sum_{j=1}^{2} ( |A,v_{k}\rangle + |\psi_{k}\rangle
+ t_{1,k}^{(1)} |u_{k},B\rangle )  \nonumber \\
\tau_{1}^{(1)}(z) & = & \frac{z^{2}}{\sqrt{2}} \left( t_{1,1}^{(1)}(z) + t_{1,2}^{(1)}(z) \right)
\nonumber \\
\tau_{2}^{(1)}(z) & = & \frac{z^{2}}{\sqrt{2}} \left( t_{1,1}^{(1)}(z) - t_{1,2}^{(1)}(z) \right)  .
\end{eqnarray}
We can now clearly see that if the graphs $G_{1}$ and $G_{2}$ are the same, which implies
that $ t_{1,1}^{(1)}(z) = t_{1,2}^{(1)}(z)$, then $\tau_{2}^{(1)}(z) = 0$.  Therefore, in this case
the particle will never enter the tail $Y_{2B}$.

We have shown that we can compute the transmission amplitudes for the combined graph,
$\Gamma$, in terms of those of $G_{1}$ and $G_{2}$, and that one of these amplitudes
vanishes if  $G_{1}$ and $G_{2}$ are the same.  Therefore, a quantum walk on the combined
graph can be used to compare $G_{1}$ and $G_{2}$.  An important issue is how efficient this
procedure is.  In particular, we would like to know how many steps the walk needs to make to
determine whether the graphs are the same or different, and how the number of steps is related
to the number of edges in $G_{1}$ and $G_{2}$.  This will remain for future work.

\section{Conclusion}
We have presented a formalism that describes quantum walks on directed, Eulerian graphs.
These graphs have incoming and outgoing tails, and we define a free quantum structure on
them that determines the motion of a particle on the graph.  We found that the transmission
amplitudes of such a graph are very useful in determining the behavior of a quantum
walk on that graph.  We have shown how the transmission amplitudes of altered graphs, i.e.\
graphs that are formed from an original one by adding or cutting an edge, can be found from
the transmission amplitudes of the original graph.  This allowed us, in some cases to find the
transmission amplitudes of a graph in terms of those of its subgraphs.  Finally, we showed
how these constructions can be used to compare two graphs by constructing a larger graph
from two smaller ones, which acts as a kind of interferometer.  If the two smaller graphs are
identical, a particle making a quantum walk on the larger graph can only emerge onto one
of the two exit tails, but not onto the other.


\begin{thebibliography}{99}
\bibitem{aharanov} Y.\ Aharanov, L,\ Davidovich, and N.\ Zagury, Phys.\ Rev.\ A {\bf 48}, 1687
(1993).
\bibitem{farhi} E.\ Farhi and S.\ Gutmann, Phys.\ Rev.\ A {\bf 58}, 915 (1998).
\bibitem{watrous}J.\ Watrous, Proceedings of the 33rd Symposium on the Theory of Computing
(STOC01) (ACM Press, New York, 2001), p.\ 60.
\bibitem{vazirani} D.\ Aharanov, A.\ Ambainis, J.\ Kempe, and U.\ Vazirani, Proceedings of the
33rd Symposium on the Theory of Computing (STOC01) (ACM Press, New York, 2001), p.\ 50,
and quant-ph/0012090.
\bibitem{shenvi} Neil Shenvi, Julia Kempe, and K.\ Birgitta Whaley, Phys.\ Rev.\ A {\bf 67},
052307 (2003).
\bibitem{ambainis1} Andris Ambainis, quant-ph/031101.
\bibitem{childs1} Andrew M.\ Childs and Jason M.\ Eisenberg, Quantum Information and
Computation {\bf 5}, 593 (2005).
\bibitem{childs2} A.\ Childs, R.\ Cleve, E.\ Deotto, E.\ Farhi, S.\ Gutman and D.\ Spielman,
Proceedings of the 35th Symposium on the Theory of Computing (STOC03) (ACM Press,
New York, 2003), p.\ 59, and quant-ph/0209131.
\bibitem{farhi2} E.\ Farhi, J.\ Goldstone, and S.\ Gutmann, quant-ph/0702144
\bibitem{childs3} Andrew M.\ Childs, Ben W.\ Reichardt, Robert Spalek, and Shengyu
Zhang, quant-ph/0703015.
\bibitem{kempe} Julia Kempe, Contemporary Physics {\bf 44}, 307 (2003).
\bibitem{kendon} Viv Kendon, quant-ph/0606016.
\bibitem{hillery} Mark Hillery, Janos Bergou, and Edgar Feldman, Phys.\ Rev.\ A {\bf 68},
032314 (2003).
\bibitem{feldman} Edgar Feldman and Mark Hillery in Coding Theory and Quantum Computing
edited by D.\ Evans, J.\ Holt, C.\ Jones, K.\ Klintworth, B.\ Parshall, O.\ Pfister, and H.\ Ward,
Contemporary Mathematics {\bf 381}, 71 (2005), and quant-ph/0403066.
\bibitem{kato} T. Kato, \emph{Perturbation Theory for Linear Operators} (Springer-Verlag,
New York, 1996), chapter 1.

\end{thebibliography}
\end{document}